\documentclass[nofootinbib,prd,twocolumn,showpacs,showkeys,preprintnumbers]{revtex4-1}
\usepackage{hyperref,amssymb,amsmath,mathrsfs,bm,graphicx}
\begin{document}

\title {Gravitational radiation within its source}%author{L. Herrera}
\author{L. Herrera}
\email{lherrera@usal.es}
\affiliation{Escuela de F\'\i sica, Facultad de Ciencias, Universidad Central de Venezuela, Caracas, Venezuela and Instituto Universitario de F\'isica
Fundamental y Matem\'aticas, Universidad de Salamanca, Salamanca, Spain}

\begin{abstract}
We review a recently proposed framework for studying axially symmetric dissipative fluids \cite{Ref1}. Some general results are discussed at the most general level. 
We then proceed to analyze some particular cases. First,  the shear-free case is considered \cite{3}.
We shall next discuss  the  perfect fluid case under the geodesic condition, without impossing ab initio the shear--free condition \cite{2}. Finally  a dissipative, geodesic fluid \cite{4}, is analyzed in some detail.
We conclude by  bringing out the attention to some open issues.
\end{abstract}
\maketitle
Talk given at the International Conference on Relativistic Astrophysics (through Skype), Lahore, February 2015. To appear in the proceedings, to be published by Punjab University Press. 9 pages, Revtex-4.
\section{INTRODUCTION}

The main purpose of the line of work outlined in this conference, is to establish the relationship between gravitational radiation and source properties. Thus, for example, we known that gravitational radiation is an irreversible process, accordingly there must exist an entropy production factor in the equation of state (dissipation) of the source.

Since we are dealing  with gravitational radiation,  we need to depart from  the spherical symmetry.  On the other hand, we shall rule out cylindrical symmetry  on physical grounds. Thus we are left with  axial and reflection symmetry, which as shown  in \cite{5bis}  is the highest degree of symmetry  of the Bondi  metric \cite{5}, which do  not prevent the emission of gravitational radiation.

We are using the $1+3$ formalism \cite{6,7,8}, in a given coordinate system, and we are going to ressort to a set of scalar functions known as Structure Scalars \cite{9}, which have been shown to be very useful in the description of self--gravitating systems \cite{10,11,12,13,14,15,16,17,17b,18b}.

\section{BASIC EQUATIONS, CONVENTIONS AND NOTATION}

We shall  consider fluid distributions endowed with axial and reflection symmetry, and we shall assume the line element to be of the form:
 \begin{equation}
ds^2=-A^2 dt^2 + B^2 \left(dr^2
+r^2d\theta^2\right)+C^2d\phi^2+2Gd\theta dt, \label{1b}
\end{equation}
 where $A, B, C, G$ are positive functions of $t$, $r$ and $\theta$, and coordinates are numbered as: $x^0=t, x^1=r, x^2= \theta, x^3=\phi$.

The energy momentum tensor describes a dissipative fluid distribution  and in its canonical form may be written as: \begin{equation}
{T}_{\alpha\beta}= (\mu+P) V_\alpha V_\beta+P g _{\alpha \beta} +\Pi_{\alpha \beta}+q_\alpha V_\beta+q_\beta V_\alpha.
\label{6bis}
\end{equation}
with
\begin{equation}
\mu = T_{\alpha \beta} V^\alpha V^\beta, \quad q_\alpha = -\mu V_\alpha - T_{\alpha \beta}V^\beta,\
\label{jc10}
\end{equation} 

\begin{equation}
P = \frac{1}{3} h^{\alpha \beta} T_{\alpha \beta},\quad   \Pi_{\alpha \beta} = h_\alpha^\mu h_\beta^\nu \left(T_{\mu\nu} - P h_{\mu\nu}\right), 
\label{jc11}  
\end{equation}
\begin{equation}
h_{\mu \nu}=g_{\mu\nu}+V_\nu V_\mu,
 \end{equation}
\begin{equation}
V^\alpha =(\frac{1}{A},0,0,0); \quad  V_\alpha=(-A,0,\frac{G}{A},0).
\label{m1}
\end{equation}
where $\mu, P, \Pi_{\alpha \beta}, q_\alpha, V_\alpha$ denote the energy density , the isotropic pressure, the anisotropic tensor, the dissipative flux and the four velocity respectively.

Next,in order to form an orthogonal tetrad,  let us  introduce the unit, spacelike vectors $\mathbf K, \mathbf L$, $\mathbf S$, with components
\begin{equation}
K_\alpha=(0,B,0,0); \quad  L_\alpha=(0,0,\frac{\sqrt{A^2B^2r^2+G^2}}{A},0),
\label{7}
\end{equation}
\begin{equation}
S_\alpha=(0,0,0,C),
\label{3n}
\end{equation}
satisfying  the following relations:
\begin{equation}
V_{\alpha} V^{\alpha}=-K^{\alpha} K_{\alpha}=-L^{\alpha} L_{\alpha}=-S^{\alpha} S_{\alpha}=-1,
\label{4n}
\end{equation}
\begin{eqnarray}
&&V_{\alpha} K^{\alpha}=V^{\alpha} L_{\alpha}=V^{\alpha} S_{\alpha}=\nonumber \\&&K^{\alpha} L_{\alpha}=K^{\alpha} S_{\alpha}=S^{\alpha} L_{\alpha}=0.
\label{5n}
\end{eqnarray}
In terms of the above vectors, the anisotropic tensor may be written as
\begin{eqnarray}
\Pi_{\alpha \beta}&=&\frac{1}{3}(2\Pi_I+\Pi_{II})(K_\alpha K_\beta-\frac{h_{\alpha
\beta}}{3})\nonumber \\&+&\frac{1}{3}(2\Pi _{II}+\Pi_I)(L_\alpha L_\beta-\frac{h_{\alpha
\beta}}{3})\nonumber \\&+&2\Pi _{KL}K_{(\alpha}L_{\beta)} \label{6bb},
\end{eqnarray}

with
\begin{eqnarray}
 \Pi _{KL}=K^\alpha L^\beta T_{\alpha \beta} 
, \quad , \label{7P}
\end{eqnarray}

\begin{equation}
\Pi_I=(2K^{\alpha} K^{\beta} -L^{\alpha} L^{\beta}-S^{\alpha} S^{\beta}) T_{\alpha \beta},
\label{2n}
\end{equation}
\begin{equation}
\Pi_{II}=(2L^{\alpha} L^{\beta} -S^{\alpha} S^{\beta}-K^{\alpha} K^{\beta}) T_{\alpha \beta}.
\label{2nbis}
\end{equation}
For the heat flux vector we may write
\begin{equation}
q_\mu=q_IK_\mu+q_{II} L_\mu,
\label{qn1}
\end{equation}
or
\begin{equation}
q^\mu=(\frac{q_{II} G}{A \sqrt{A^2B^2r^2+G^2}}, \frac{q_I}{B}, \frac{Aq_{II}}{\sqrt{A^2B^2r^2+G^2}}, 0),\label{q}
\end{equation}
\begin{equation}
 q_\mu=\left(0, B q_I, \frac{\sqrt{A^2B^2r^2+G^2}q_{II}}{A}, 0\right).
\label{qn}
\end{equation}

\subsection{Kinematical variables}

The kinematical variables (the four acceleration, the expansion, the shear tensor and the vorticity) are defined respectively as:
\begin{eqnarray}
a_\alpha&=&V^\beta V_{\alpha;\beta}=a_I K_\alpha+a_{II}L_\alpha\nonumber\\
&=&\left(0, \frac {A_{,r} }{A },\frac{G}{A^2}\left[-\frac {A_{,t}}{A}+\frac {G_{,t}}{G}\right]+\frac {A_{,\theta}} {A},0\right),
\label{acc}
\end{eqnarray}
\begin{eqnarray}
\Theta&=&V^\alpha_{;\alpha}\nonumber\\
&=&\frac{AB^2}{r^2A^2B^2+G^2}\,\left[r^2\left(2\frac{B_{,t}}{B}+\frac{C_{,t}}{C}\right)\right.\nonumber\\
&&+\left.\frac{G^2}{A^2B^2}\left(\frac{B_{,t}}{B}-\frac{A_{,t}}{A}+\frac{G_{,t}}{G}+\frac{C_{,t}}{C}\right)\right],
\label{theta}
\end{eqnarray}

\begin{equation}
\sigma_{\alpha \beta}= V_{(\alpha;\beta)}+a_{(\alpha}
V_{\beta)}-\frac{1}{3}\Theta h_{\alpha \beta}, \label{bacc}
\end{equation}
or

\begin{eqnarray}
\sigma _{\alpha \beta}=\frac{1}{3}(2\sigma _I+\sigma_{II}) (K_\alpha
K_\beta-\frac{1}{3}h_{\alpha \beta})\nonumber \\+\frac{1}{3}(2\sigma _{II}+\sigma_I) (L_\alpha
L_\beta-\frac{1}{3}h_{\alpha \beta}),\label{sigmaT}
\end{eqnarray}
where
\begin{eqnarray}
2\sigma _I+\sigma_{II}&=&\frac{3}{A}\left(\frac{B_{,t}}{B}-\frac{C_{,t}}{C}\right), \label{sigmasI}
\end{eqnarray}
\begin{eqnarray}
2\sigma _{II}+\sigma_I&=&\frac{3}{A^2B^2r^2+G^2}\,\left[AB^2r^2\left(\frac{B_{,t}}{B}-\frac{C_{,t}}{C}\right)\right.\nonumber\\
&
&\left.+\frac{G^2}{A}\left(-\frac{A_{,t}}{A}+\frac{G_{,t}}{G}-\frac{C_{,t}}{C}\right)\right] \label{sigmas},
\end{eqnarray}

\begin{equation}
\omega_\alpha=\frac{1}{2}\,\eta_{\alpha\beta\mu\nu}\,V^{\beta;\mu}\,V^\nu=\frac{1}{2}\,\eta_{\alpha\beta\mu\nu}\,\Omega
^{\beta\mu}\,V^\nu,\label{vomega}
\end{equation}
where $\Omega_{\alpha\beta}=V_{[\alpha;\beta]}+a_{[\alpha}
V_{\beta]}$, $\omega_\alpha$ and $\eta_{\alpha\beta\mu\nu}$ denote the vorticity tensor, the vorticity vector  and the Levi-Civita tensor, respectively;

\begin{equation}
\Omega_{\alpha\beta}=\Omega (L_\alpha K_\beta -L_\beta
K_{\alpha}),\label{omegaT}
\end{equation}

\begin{equation}
\omega _\alpha =-\Omega S_\alpha.
\end{equation}

\begin{equation}
\Omega =\frac{G(\frac{G_{,r}}{G}-\frac{2A_{,r}}{A})}{2B\sqrt{A^2B^2r^2+G^2}}.
\label{no}
\end{equation}

Observe that from (\ref{no}) and regularity conditions at the centre, it follows that: $G=0\Leftrightarrow \Omega=0$.
\subsection{The orthogonal splitting of the Riemann  Tensor and structure scalars}
Using the well kown decomposition of the Riemann tensor in terms of the Weyl tensor, the Ricci tensor and the Ricci scalar, and linking the two later variables with the energy momentum tensor, via the Einstein equations, it can be shown that the Riemann tensor may be written as:
\begin{equation}
R^{\alpha \beta}_{\quad \nu \delta}=R^{\alpha \beta}_{(F)\,\,
\nu\delta}+R^{\alpha \beta}_{(Q)\,\, \nu \delta}+R^{\alpha
\beta}_{(E)\,\, \nu\delta}+R^{\alpha \beta}_{(H)\,\, \nu \delta},
\label{dc1}
\end{equation}
 with
\begin{eqnarray}
R^{\alpha \beta}_{(F)\,\,
\nu\delta}=\frac{16\pi}{3}(\mu+3P)V^{[\alpha}V_{[\nu}
h^{\beta]}_{\delta ]}+\frac{16\pi}{3}\mu h^\alpha_{[\nu}
h^\beta_{\delta ]}, \label{dc2}
\end{eqnarray}
%\begin{widetext}
\begin{eqnarray}
R^{\alpha \beta}_{(Q)\,\, \nu \delta}=-16\pi V^{[\alpha
}h^{\beta]}_{[\nu}q_{\delta ]}-16\pi  V_{[\nu}h ^{[\alpha
}_{\delta]}q^{\beta]}-\\ \nonumber -16\pi V ^{[\alpha}V_{[ \nu}\Pi
^{\beta]}_{\delta]}+ 16\pi h^{[\alpha}_{[\nu}\Pi ^{\beta
]}_{\delta]}\label{dc3}
\end{eqnarray}
%\end{widetext}
\begin{eqnarray}
 R^{\alpha \beta}_{(E)\,\, \nu\delta}&=&4V^{[\alpha}V_{[\nu} E^{\beta]}_{\delta ]}+
4 h^{[\alpha}_{[\nu}E^{\beta ]}_{\delta ]}, \label{dc4}
\end{eqnarray}
\begin{eqnarray}
 R^{\alpha\beta}_{(H)\,\, \nu \delta}&=&-2 \epsilon ^{\alpha \beta
 \gamma}V_{[\nu}H_{\delta]\gamma}-2\epsilon _{\nu\delta\gamma}V^{[\alpha}H^{\beta
 ]\gamma}.\label{dc5}
\end{eqnarray}
In the above,  $E_{\alpha \beta}, H_{\alpha\beta}$ denote the electric and magnetic parts of the Weyl tensor, respectively, defined as usual by:
\begin{equation}
E_{\alpha \beta}=C_{\alpha\nu\beta\delta}V^\nu V^\delta,\qquad
H_{\alpha\beta}=\frac{1}{2}\eta_{\alpha \nu \epsilon
\rho}C^{\quad \epsilon\rho}_{\beta \delta}V^\nu
V^\delta\,,\label{EH}
\end{equation}

\noindent where $\epsilon _{\alpha \beta \rho}=\eta_{\nu
\alpha \beta \rho}V^\nu$.

In our case these tensors may be written in terms of five scalar functions as:
%\begin{widetext}
\begin{eqnarray}
E_{\alpha\beta}&=&\frac{1}{3}(2\mathcal{E}_I+\mathcal{E}_{II}) (K_\alpha
K_\beta-\frac{1}{3}h_{\alpha \beta})\nonumber \\&+&\frac{1}{3}(2\mathcal{E}_{II}+\mathcal{E}_{I}) (L_\alpha
L_\beta-\frac{1}{3}h_{\alpha \beta})\nonumber \\
&+&\mathcal{E}_{KL} (K_\alpha
L_\beta+K_\beta L_\alpha), \label{E'}
\end{eqnarray}
%\end{widetext}
\noindent

\begin{equation}
H_{\alpha\beta}=H_1(S_\alpha K_\beta+S_\beta
K_\alpha)+H_2(S_\alpha L_\beta+S_\beta L_\alpha)\label{H'}.
\end{equation}
Let us now introduce the following tensors
\begin{equation}
Y_{\alpha \beta}=R_{\alpha \nu \beta \delta}V^\nu V^\delta,
\label{Y}
\end{equation}
\begin{equation}
X_{\alpha \beta}=\frac{1}{2}\eta_{\alpha\nu}^{\quad \epsilon
\rho}R^\star_{\epsilon \rho \beta \delta}V^\nu V^\delta,\label{X}
\end{equation}

\begin{equation}
Z_{\alpha\beta}=\frac{1}{2}\epsilon_{\alpha \epsilon \rho}R^{\quad
\epsilon\rho}_{ \delta \beta} V^\delta,\label{Z}
\end{equation}
 where $R^\star _{\alpha \beta \nu
\delta}=\frac{1}{2}\eta_{\epsilon\rho\nu\delta}R_{\alpha
\beta}^{\quad \epsilon \rho}$.

Or, using (\ref{dc1})
\begin{eqnarray}
Y_{\alpha \beta}&=&\frac{1}{3}Y_T h_{\alpha \beta} + \frac{1}{3}(2Y_I+Y_{II}) (K_\alpha
K_\beta-\frac{1}{3}h_{\alpha \beta})\nonumber \\
&+&\frac{1}{3}(2Y_{II}+Y_{I}) (L_\alpha L_\beta-\frac{1}{3}h_{\alpha \beta})\nonumber \\&+&Y_{KL}
(K_\alpha L_\beta+K_\beta L_\alpha), \label{yf}
\end{eqnarray}
with
\begin{eqnarray}
Y_T=4\pi(\mu+3P), \label{ortc1}\\
Y_I=\mathcal{E}_I-4\pi \Pi_I, \label{ortc2}\\
Y_{II}=\mathcal{E}_{II}-4\pi \Pi_{II}, \label{YY}\\
Y_{KL}=\mathcal{E}_{KL}-4\pi \Pi_{KL}.\label{KL}
\end{eqnarray}

\begin{eqnarray}
X_{\alpha \beta}&=&\frac{1}{3}X_T h_{\alpha \beta} +  \frac{1}{3}(2X_I+X_{II}) (K_\alpha
K_\beta-\frac{1}{3}h_{\alpha \beta})\nonumber \\
&+&\frac{1}{3}(2X_{II}+X_{I}) (L_\alpha L_\beta-\frac{1}{3}h_{\alpha \beta})\nonumber \\&+&X_{KL}
(K_\alpha L_\beta+K_\beta L_\alpha), \label{xf}
\end{eqnarray}
with
\begin{eqnarray}
X_T=8\pi \mu, \label{ortc1x}\\
X_I=-\mathcal{E}_I-4\pi \Pi_I, \label{ortc2x}\\
X_{II}=-\mathcal{E}_{II}-4\pi \Pi_{II}, \label{YYx}\\
X_{KL}=-\mathcal{E}_{KL}-4\pi \Pi_{KL}.\label{KLx}
\end{eqnarray}
Finally
\begin{equation}
Z_{\alpha\beta}=H_{\alpha\beta}+4\pi q^\rho \epsilon_{\alpha\beta
\rho}. \label{Z'}
\end{equation}

or

\begin{equation}
Z_{\alpha\beta}=Z_IK_\beta S_\alpha+Z_{II}K_\alpha S_\beta+Z_{III}L_\alpha S_\beta+Z_{IV}L_\beta S_\alpha \label{Z2}
\end{equation}
where 

\begin{eqnarray}
Z_I&=&(H_1-4\pi q_{II});\quad Z_{II}=(H_1+4\pi  q_{II}); \nonumber \\Z_{III}&=&(H_2-4\pi q_I); \quad  Z_{IV}=(H_2+4\pi q_I). \label{Z2}
\end{eqnarray}
Variables:$Y_{T, I, II, KL}, X_{T, I, II, KL}, Z_{I, II, III, IV}$ are the structure scalars of our distribution.
\subsection{The super--Poynting
vector}
An important role in our discussion is played by the super--poynting vector. Indeed, we recall that  we define a state of intrinsic gravitational radiation (at any given point), to be one in which  the super-Poynting vector does not vanish for any  unit timelike vector   \cite{18,19,20}. Then since the vanishing of the magnetic part of the Weyl tensor implies the vanishing of the super-Poynting vector, it is clear that  FRW does not produce gravitational radiation. It is also  worth recalling that the tight link between the super-Poynting vector and the existence of a state of radiation, is firmly supported by the relationship between the former and the Bondi news function \cite{5,21} (see \cite{22} for a discussion on this point).

Then from the definition of the super-Poynting vector,
\begin{equation}
P_\alpha = \epsilon_{\alpha \beta \gamma}\left(Y^\gamma_\delta
Z^{\beta \delta} - X^\gamma_\delta Z^{\delta\beta}\right), 
\label{SPdef}
\end{equation}
we obtain
\begin{equation}
  P_\alpha=P_I K_\alpha+P_{II} L_\alpha,
\label{nsp}
\end{equation}
with
%\begin{widetext}
\begin{eqnarray}
P_I &=
&\frac{H_2}{3}(2Y_{II}+Y_I-2X_{II}-X_I)+H_1(Y_{KL}-X_{KL})\nonumber \\ &+& \frac{4\pi q_I}{3}\left[2Y_T+2 X_T-X_I-Y_I\right] 
- 4\pi q_{II}(X_{KL} +Y_{KL}),\nonumber
\\
P_{II}&=&\frac{H_1}{3}(2X_{I}+X_{II}-Y_{II}-2Y_I)+H_2(X_{KL}-Y_{KL}) \nonumber \\&-&4\pi q_I(Y_{KL}+X_{KL})
\nonumber \\&+&\frac{4\pi q_{II}}{3}\left[2Y_T+2X_T-X_{II}-Y_{II}\right]. \label{SPP}
\end{eqnarray}
%\end{widetext}

  Both components  have terms not containing heat dissipative contributions. It is reasonable to associate these with gravitational radiation.   Also, note that both components  have contributions of  both components of the heat flux vector.

There is always a non-vanishing component of $P^\mu$, on the
plane orthogonal to a unit vector along which there is a non-vanishing component of vorticity (the $\theta-r$- plane).
Inversely, $P^\mu$ vanishes along the $\phi$-direction since there are no motions along this latter direction, because of the reflection symmetry. 

 We can identify three different contributions in (\ref{SPP}). On the one hand we have contributions from the  heat transport process. These are independent of the magnetic part of the Weyl tensor, which explains why they  remain in the spherically symmetric limit. 

On the other hand  we have contributions from the magnetic part of the Weyl tensor. These are of two kinds: a)  contributions associated with the propagation of gravitational radiation within the fluid, b) contributions of the flow of super--energy associated with the vorticity on the plane orthogonal to the direction of propagation of the radiation. Both  are intertwined, and it appears  to be impossible to disentangle them  through two independent scalars.

As mentioned before, both components of the heat flux four-vector, appear  in both components of the super--Poynting vector. Observe that this is achieved  through the $X_{KL}+Y_{KL}$ terms in (\ref{SPP}), or using (\ref{KL}, \ref{KLx}), through $\Pi_{KL}$.  Thus,    $\Pi_{KL}$ couples the two components of the super--Poynting vector, with the two components of the heat flux vector.
\section{THE EQUATIONS}
We shall  now deploy the whole set of equations for the variables defined so far.
\subsection{The heat transport equation}

We shall need a transport equation derived from  a causal  dissipative theory (e.g. the
M\"{u}ller-Israel-Stewart second
order phenomenological theory for dissipative fluids \cite{23,24,25,26}).

Indeed,  the Maxwell-Fourier law for
radiation flux leads to a parabolic equation (diffusion equation)
which predicts propagation of perturbations with infinite speed
(see \cite{27}-\cite{29} and references therein). This simple fact
is at the origin of the pathologies \cite{30} found in the
approaches of Eckart \cite{31} and Landau \cite{32} for
relativistic dissipative processes. To overcome such difficulties,
various relativistic
theories with non-vanishing relaxation times have been proposed in
the past \cite{23,24,25,26,33,34}. The important point is that
all these theories provide a heat transport equation which is not
of Maxwell-Fourier type but of Cattaneo type \cite{35}, leading
thereby to a hyperbolic equation for the propagation of thermal
perturbations.

A fundamental parameter   in these theories is the relaxation time $\tau$ of the
corresponding  dissipative process. This positive--definite quantity has a
distinct physical meaning, namely the time taken by the system to return
spontaneously to the steady state (whether of thermodynamic equilibrium or
not) after it has been suddenly removed from it. 
Therefore, when studying transient regimes, i.e., the evolution between two
steady--state situations,  $\tau$ cannot be neglected. In 
fact, leaving aside that parabolic theories are necessarily non--causal,
it is obvious that whenever the time scale of the problem under
consideration becomes of the order of (or smaller than) the relaxation time,
the latter cannot be ignored, since 
neglecting the relaxation time amounts -in this situation- to
disregarding the whole problem under consideration.

Thus, the transport equation for the heat flux reads \cite{24,25,28},
\begin{equation}
\tau h^\mu_\nu q^\nu _{;\beta}V^\beta +q^\mu=-\kappa
h^{\mu\nu}(T_{,\nu}+T a_\nu)-\frac{1}{2}\kappa T^2\left
(\frac{\tau V^\alpha}{\kappa T^2}\right )_{;\alpha}q^\mu,\label{qT}
\end{equation}

\noindent where $\tau$, $\kappa$, $T$ denote the relaxation time,
the thermal conductivity and the temperature, respectively.

Contracting (\ref{qT}) with $L_\mu$ we obtain

\begin{eqnarray}
\frac{\tau}{A}\left(q_{II,t}+A q_{I} \Omega\right)+q_{II}&=&-\frac{\kappa}{A}\left(\frac{G T_{,t}+A^2 T_{,\theta}}{\sqrt{A^2B^2r^2+G^2}}+A T a_{II}\right)\nonumber \\& -&\frac{\kappa T^2q_{II}}{2}(\frac{\tau V^\alpha}{\kappa T^2})_{;\alpha},\label{qT1n}
\end{eqnarray}

where  (\ref{no}), has been used.

On other hand,  contracting (\ref{qT}) with $K_\mu$, we find

\begin{eqnarray}
\frac{\tau}{A}\left(q_{I,t}-A q_{II} \Omega\right)+q_{I}=-\frac{\kappa}{B}(T_{,r}+BTa_I)\nonumber \\
-\frac{\kappa T^2 q_{I}}{2}(\frac{\tau
V^\alpha}{\kappa T^2})_{;\alpha}. \label{qT2n}
\end{eqnarray}

 It is worth noting  that the two equations above are coupled  through the vorticity. 
This fact entails an interesting thermodynamic consequence.
Indeed, let us assume that at some initial time (say $t=0$) and before it, there is thermodynamic equilibrium  in the  $\theta$ direction, this implies $q_{II}=0$, and  also that the  corresponding Tolman's temperature  \cite{36}  is constant, which in turns implies that the term within the round bracket in the first term on the right of (\ref{qT1n}) vanishes. Then it follows at once from (\ref{qT1n}) that:
\begin{equation}
q_{II,t}=-A\Omega q_I,
\label{nvz1}
\end{equation}
implying that the propagation in time of the vanishing of the meridional flow, is subject to the vanishing of the vorticity and/or the vanishing of  heat flow in the $r$- direction.

Inversely, repeating the same argument for (\ref{qT2n}) we obtain at the initial time when we assume thermodynamic equilibrium,

\begin{equation}
q_{I,t}=A\Omega q_{II}.
\label{nvz2}
\end{equation}

Thus, it appears that the vanishing of the radial component of the heat flux vector at some initial time, will propagate in time  if only, the vorticity and/or  the meridional heat flow vanish.

In other words, time propagation of the  thermal equilibrium condition, in either direction $r$ or $\theta$, is assured  only in the absence of vorticity. Otherwise, it requires initial thermal  equilibrium in both directions.

This result is a clear reminiscence of the von Zeipel's theorem \cite{37}.
\subsection{The equations for the metric functions, the kinematical  variables and the Riemann tensor components.}
Let us first recall the decomposition of the covariant derivative of the four--velocity in terms of the kinematical variables given by:
\begin{equation}
V_{\alpha;\beta}=\sigma_{\alpha \beta}+\Omega_{\alpha \beta}-a_\alpha V_\beta+\frac{1}{3}h_{\alpha \beta}\Theta,
\label{conf1}
\end{equation}
which entails all the equations (\ref{acc}), (\ref{theta}), (\ref{bacc}), (\ref{vomega}).

Now, if we regard the above expression as a first order differential equation relating the kinematical variables with first order derivative of the metric functions, and look for its integrability conditions, we find
\begin{equation}
V_{\alpha ;\beta ; \nu}- V_{\alpha;\nu;\beta}=R^{\mu}_{\alpha
\beta \nu}V_\mu.
\label{conf2}
\end{equation}

From this last equation the following equations are obtained, by projecting with different combinations of the tetrad vectors:

An evolution equation for the expansion scalar (the Raychaudhuri equation)
\begin{equation}
\Theta _{;\alpha}V^\alpha +\frac{1}{3}\Theta ^2+2(\sigma ^2-\Omega
^2)-a^\alpha _{;\alpha}+4\pi(\mu+3P)=0\label{ec3}
\end{equation}

 where $2\sigma ^2=\sigma _{\alpha\beta} \sigma
^{\alpha\beta}$.

An equation for the evolution of the shear tensor:
%\begin{widetext}
\begin{eqnarray}
&&h^\mu_{\alpha}h^\nu_{\beta}\sigma_{\mu\nu;\delta}V^\delta+\sigma_\alpha
^\mu \sigma_{\beta \mu}+\frac{2}{3}\Theta \sigma_{\alpha
\beta}\nonumber \\&-&\frac{1}{3}\left ( 2\sigma ^2+\Omega^2-a^\delta
_{;\delta}\right) h_{\alpha \beta}
+\omega _\alpha \omega _\beta-a_\alpha a_\beta\nonumber \\&-&h^\mu
_{(\alpha}h^\nu_{\beta)}a_{\nu;\mu}+E_{\alpha \beta}-4\pi
\Pi_{\alpha \beta}=0.\label{ec4}
\end{eqnarray}
%\end{widetext}

An equation for the evolution of the vorticity tensor:
\begin{equation}
 h^\mu _{\alpha}h^\nu _{\beta}\Omega _{\mu\nu;\delta}V^\delta
+\frac{2}{3}\Theta \Omega _{\alpha\beta}
+2\sigma_{\mu[\alpha}\Omega ^\mu
_{\,\,\,\beta]}-h^\mu_{[\alpha}h^\nu
_{\beta]}a_{\mu;\nu}=0.\label{ec51}
\end{equation}

Two constraint equations relating the kinematical variables and their derivatives with the heat flux vector and the magnetic part of the Weyl tensor:
\begin{equation}
h^\beta_\alpha \left (\frac{2}{3}\Theta_{;\beta}-\sigma ^\mu
_{\beta;\mu}+\Omega ^{\,\,\,\mu} _{\beta \,\,\,;\mu}\right )+\left
(\sigma_{\alpha\beta}+\Omega _{\alpha \beta}\right )a^\beta=8\pi
q_\alpha, \label{ec61}
\end{equation}

\begin{equation}
2\omega _{(\alpha }a_{\beta)}+h^\mu _{(\alpha}h_{\beta )\nu}\left
( \sigma_{\mu \delta}+\Omega _{\mu \delta}\right
)_{;\gamma}\eta^{\nu\kappa\gamma\delta}V_\kappa=H_{\alpha
\beta}.\label{ec6}
\end{equation}
\subsection{The conservation equations}

The conservation law $T^\alpha _{\beta;\alpha}=0$, leads to the following equations:
%\begin{widetext}
\begin{eqnarray}
\mu _{;\alpha}V^\alpha +(\mu+P)\Theta +\frac{1}{9}(2\sigma _{I}+\sigma _{II})\Pi_I\nonumber \\+\frac{1}{9}(2\sigma_{II}+\sigma _I)\Pi_{II}+ q^{\alpha}_{;\alpha} + q^\alpha a_\alpha =0,\label{esc1}
\end{eqnarray}

\begin{eqnarray}
(\mu+P)a_\alpha+h_{\alpha}^\beta\left
(P_{;\beta}+\Pi_{\beta;\mu}^\mu+q_{\beta;\mu}V^\mu\right )\nonumber \\+\left(
\frac{4}{3}\Theta h_{\alpha \beta}+\sigma_{\alpha
\beta}+\Omega_{\alpha\beta}\right )q^\beta=0.\label{ec2}
\end{eqnarray}
\subsection{The Bianchi identities}
Next, if we regard (\ref{conf2}) as a system of differential equations of first order, relating the Riemann tensor components with the kinematical variables and their derivatives, and look for their integrability conditions, we are lead to the Bianchi idenitities, which together with (\ref{dc1}), lead to the following set of equations:

An evolution equation for the electric part of the Weyl tensor
%\begin{widetext}
\begin{eqnarray}
&&h^\mu_{(\alpha} h^\nu _{\beta)} E_{\mu\nu;\delta}V^\delta +\Theta
E_{\alpha\beta}+h_{\alpha\beta}E_{\mu\nu}\sigma
^{\mu\nu}-3E_{\mu(\alpha}\sigma ^\mu _{\beta )}\nonumber \\&+&h^\mu
_{(\alpha}\eta _{\beta )}^{\,\,\,\, \delta \gamma\kappa}V_\delta
H_{\gamma\mu;\kappa}
-E_{\delta
(\alpha}\Omega_{\beta)}^{\,\,\,\delta}
-2H^\mu _{(\alpha}\eta_{\beta)\delta \kappa \mu }V^\delta
a^\kappa=\nonumber \\&-&4\pi(\mu+P)\sigma _{\alpha \beta}-\frac{4\pi}{3}\Theta
\Pi_{\alpha \beta}-4\pi h^\mu_{(\alpha} h^\nu_{\beta)}
\Pi_{\mu\nu;\delta}V^\delta-4\pi\sigma_{\mu(\alpha}\Pi_{\beta)}^\mu
\nonumber \\&-&4\pi\Omega ^\mu_{\,\,\,(\alpha}\Pi_{\beta )\mu}  8\pi
a_{(\alpha}q_{\beta )}+\frac{4\pi}{3}\left
(\Pi_{\mu\nu}\sigma^{\mu\nu}+a_\mu q^\mu+q^\mu _{;\mu}\right
)h_{\alpha\beta}\nonumber \\&-&4\pi h^\mu_{(\alpha}h_{\beta )}^\nu
q_{\nu;\mu}.\label{ec7}
\end{eqnarray}
A constraint equation for the spatial derivatives of the electric part of the Weyl tensor
%\end{widetext}
%\begin{widetext}
\begin{eqnarray}
&&h^\mu _\alpha h^{\nu
\beta}E_{\mu\nu;\beta}-\eta_{\alpha}^{\,\,\,\delta \nu
\kappa}V_\delta \sigma ^\gamma _\nu
H_{\kappa\gamma}+3H_{\alpha\beta}\omega ^\beta=
\frac{8\pi}{3}h^\beta _\alpha\mu_{;\beta}\nonumber \\&-&4\pi h^\beta_\alpha
h^{\mu\nu}\Pi_{\beta \nu;\mu}-4\pi \left ( \frac{2}{3}\Theta
h_\alpha ^\beta-\sigma ^\beta _\alpha+3\Omega ^{\,\,\,\beta}
_\alpha \right )q_\beta, \label{ec8}
\end{eqnarray}
A constraint equation for the spatial derivatives of the magnetic part of the Weyl tensor
%\end{widetext}
%\begin{widetext}
\begin{eqnarray}
&&\left ( \sigma _{\alpha \delta}E^\delta _{\beta}+3\Omega _{\alpha
\delta}E^\delta _\beta \right)\epsilon _\kappa ^{\,\,\, \alpha
\beta}+a^\nu H_{\nu\kappa}-H^{\nu\delta}_{\,\,\,\, ;\delta}h_{\nu
\kappa} =\nonumber \\&&4\pi (\mu+P)\Omega _{\alpha\beta}\epsilon _\kappa ^{\,\,\,\alpha
\beta}\nonumber \\&+&4\pi \left [q_{\alpha ;\beta}+\Pi _{\nu\alpha}(\sigma ^\nu
_{\,\,\beta}+\Omega ^\nu _{\,\,\,\beta})\right ]\epsilon _\kappa
^{\,\,\,\alpha \beta},  \label{ec9}
\end{eqnarray}
%\end{widetext}
An evolution equation for the magnetic part of the Weyl tensor

%\begin{widetext}
\begin{eqnarray}
&&2a_\beta E_{\alpha\kappa}\epsilon_{\gamma}^{\,\,\,\alpha
\beta}-E_{\nu\beta;\delta}h^\nu_{\kappa}\epsilon_{\gamma}^{\,\,\,
\delta \beta}+E^\delta_{\beta;\delta}\epsilon_{\gamma \kappa}^{\quad
\beta}+\frac{2}{3}\Theta H_{\kappa \gamma}\nonumber \\&+&H^\mu_{\nu;\delta}V^\delta
h^\nu _{\kappa}h_{\mu \gamma}
-\left (\sigma _{\kappa\delta}+\Omega_{\kappa \delta}\right
)H^\delta _\gamma\nonumber \\&+&\left (\sigma _{\beta \delta}+\Omega _{\beta
\delta}\right )H^\mu _\alpha
\epsilon_{\kappa\,\,\,\mu}^{\,\,\,\delta}\epsilon_{\gamma}^{\,\,\,\alpha\beta}
+\frac{1}{3}\Theta H^\mu _\alpha
\epsilon_{\kappa\,\,\,\mu}^{\,\,\,\delta}\epsilon_{\gamma\,\,\,\delta}^{\,\,\,\alpha}\nonumber
\\
&=&\frac{4\pi}{3}\mu_{,\beta}\epsilon_{\gamma\kappa}^{\quad \beta}+4\pi
\Pi_{\alpha\nu;\beta}h^\nu_{\kappa}\epsilon_{\gamma}^{\,\,\,\alpha
\beta}\nonumber \\&+&4\pi\left [q_\kappa\Omega_{\alpha \beta}+q_\alpha
(\sigma_{\kappa\beta}+\Omega_{\kappa\beta}+\frac{1}{3}\Theta
h_{\kappa\beta})\right]\epsilon_{\gamma}^{\,\,\,\alpha\beta}.\label{ec10}
\end{eqnarray}

Equations (\ref{qT}), (\ref{ec3})-(\ref{ec10}) form the full set of equations for the variables of our problem. However, the following remarks are in order at this point:
\begin{itemize}
\item Obviously, not all of these equations  are independent, however depending on the problem under consideration, it may be more advantageous to use one subset instead of the other, and therefore here we present them all. 
\item The scalar equations obtained by projecting  the above system, on all possible combinations of tetrad vectors, are deployed in the Appendix B of \cite{Ref1}.
\item The obtained equations are of first order, unlike the Einstein equations, which are differential equations of second order for the metric functions. This reduction is achieved by enlarging the number of variables and equations.
\item  In the case of specific modeling, another important question arises, namely: what additional information is required to close the system of equations? It is clear that  information about local physical aspects of the source (e.g. equations of state and/or information about energy production) are not included in the set of deployed equations and therefore should be given, in order that metric and matter functions could be solved for in terms of initial data. 
\end{itemize}
\section{THE EFFECTIVE INERTIAL MASS DENSITY OF THE DISSIPATIVE FLUID}
In classical dynamics the inertial mass is defined as the factor of
proportionality between the three-force applied to a particle (a fluid
element) and the resulting three-acceleration, according to Newton's
second law.

In relativistic dynamics a similar relation only holds (in general) in the
instantaneous rest frame (i.r.f.), since the three-acceleration and the
force that causes it are not (in general) paralell, except in the i.r.f.
(see for example \cite{38}).

We shall derive below, an expression for the effective inertial
mass density for our dissipative fluid distribution.

By ``effective inertial mass'' (e.i.m.) density we mean the factor of
proportionality between the applied three-force density and the resulting
proper acceleration (i.e., the three-acceleration measured in the i.r.f.).

As we shall see, the obtained expression for the e.i.m. density contains
a contribution from dissipative variables which reduces its value with
respect to the non-dissipative situation. Such decreasing of e.i.m.
density was  brought out for the first time in the spherically symmetric
self-gravitating case in \cite{39}. Afterwards this effect was also detected in  the axially symmetric
self-gravitating case \cite{40}, for slowly rotating
self-gravitating systems \cite{41}, and under other many different circumstances \cite{HS,eim,essay,42,43,44}.

 It is perhaps worth noticing that
the concept of effective inertial mass is familiar in other branches of physics, thus
for example the e.i.m. of an electron moving under a given force through
a crystal, differs from the value corresponding to an electron moving
under the same force in free space, and may even become negative (see
\cite{45,46}).

Combining the equations (\ref{ec2}) and (\ref{qT}) we obtain 

\begin{eqnarray}
&&(\mu+P)\left(1-\alpha\right)a_\alpha=-h^\beta _{\alpha}\Pi ^\mu _{\beta ;\mu}-\nabla _\alpha P\nonumber \\&-&(\sigma _{\alpha \beta}+\Omega _{\alpha \beta})q^\beta\nonumber
+\frac{\kappa}{\tau}\nabla _\alpha T\nonumber \\&+&\left \{\frac{1}{\tau}+\frac{1}{2}D_t \left[ln(\frac{\tau}{\kappa T^2})\right]-\frac{5}{6}\Theta \right\}q_\alpha,
\label{eim}
\end{eqnarray}
an expression which takes the desired, "Newtonian", form.
$$
\mbox{Force=e.i.m.}\times\mbox{acceleration(proper),}
$$
where $\nabla _\alpha P\equiv h^\beta _\alpha P_{,\beta}$,  $D_t f \equiv f_{,\beta}V^\beta$  and $\alpha=\frac{\kappa T}{\tau (\mu+P)}.$

 The factor multiplying the four acceleration vector represents the effective inertial mass density.
Thus, the obtained expression for the e.i.m. density contains a contribution from dissipative variables, which reduces its value with respect to the non-dissipative situation.

From the equivalence principle it follows that the ``passive'' gravitational mass density should be reduced too, by the same factor. This in turn might lead, in some critical cases when such diminishing is significative, to a bouncing of the  collapsing object.

It should be observed that causality and stability conditions hindering
the system to attain the condition $\alpha=1$, are obtained on the basis of a
linear approximation, whose validity, close to the critical point
($\alpha=1$), is questionable \cite{52}.

At any rate, examples of fluids attaining the critical point and
exhibiting reasonble physical properties have been presented elsewhere
\cite{53,54}.

In order to evaluate $\alpha$, let us turn back to c.g.s. units. Then,
assuming for simplicity $\mu + p \approx 2\mu$, we obtain
\begin{equation}
\frac{\kappa T}{\tau (\mu + p)} \approx \frac{[\kappa][T]}{[\tau][\mu]}
\times 10^{-42}
\label{un}
\end{equation}
where $[\kappa]$, $[T]$, $[\tau]$, $[\mu]$ denote the numerical values of
these quantities in $erg. \, s^{-1} \, cm^{-1} \, K^{-1}$, $K$, $s$ and
$g.\, cm^{-3}$, respectively.

Obviously, this will be a very small quantity (compared to $1$), unless
conditions for extremely high values of $\kappa$ and $T$ are attained.

At present we may speculate that  $\alpha$ may
increase substantially (for a non-negligible values of $\tau$) in a pre-supernovae event

Indeed, at the last stages of massive star evolution, the decreasing of the opacity
of the fluid, from very high values preventing the propagation of photons
and neutrinos (trapping \cite{55}), to smaller values, gives rise to
radiative heat conduction. Under these conditions both $\kappa$ and $T$
could be sufficiently large as to imply a substantial increase of
$\alpha$. Indeed, the values suggested in \cite{56} ($[\kappa] \approx
10^{37}$;
$[T] \approx 10^{13}$; $[\tau] \approx 10^{-4}$; $[\mu] \approx
10^{12}$ ) lead to $\alpha \approx 1$. The obvious consequence of which
would be to enhance the efficiency of whatever expansion mechanism, of
the central core, at place, because of the decreasing of its e.i.m.
density.
At this point it is worth noticing that the relevance of relaxational effects on gravitational collapse has been often  exhibited and stressed (see \cite{47,48,49,50,51}, and references therein)

It is also worth noticing that  the inflationary equation of state
(in the perfect fluid case) $\mu + P = 0$, is, as far as the equation of
motion is concerned, equivalent to $\alpha = 1$ in the dissipative case (both imply the vanishing of the e.i.m. density).

Finally, it is worth stressing that it is the first term on the left  and the second on the right, in (\ref{qT}) the direct responsible for the decreasing in the e.i.m density. Therefore any hyperbolic dissipative theory yielding a
Cattaneo-type equation in the non-relativistic limit, is expected to give a result similar to the one obtained here.
\section{SOME PARTICULAR CASES}
In what follows we shall   consider some particular cases, where some variables (e. g. the shear) are assumed  to vanish. We do so, on the one hand for simplicity, and on the other, in order to bring out the role of some specific variables. However, it should be kept in mind that such kinds of ``suppressions'' may lead to inconsistencies in the set of equations. This is for example the case of ``silent'' universes \cite{s1,s2}, where dust sources have vanishing magnetic Weyl tensor, and lead to a system of 1+3 constraint equations that do not seem to be integrable in general \cite{s3}. In other words for any specific modeling,  the possible occurrence of these types of inconsistencies should be carefully considered.
\subsection{The shear free case}
This case has been analyzed in detail in \cite{3}. Below we summarize the main results obtained under the shear--free condition.
\begin{itemize}
\item For a general dissipative and anisotropic (shear free) fluid, vanishing vorticity, is a  necessary and sufficient condition for the magnetic part of the Weyl tensor  to vanish.

\item Vorticity should necessarily appear if the system  radiates gravitationally. This further reinforces  the well established  link between radiation and vorticity.

\item In the geodesic (shear--free) case,  the vorticity  vanishes (and thereof  the magnetic part of the Weyl tensor). No gravitational radiation is produced. 
A similar result is obtained for the cylindrically symmetric case,  suggesting a link between the shear of the source and the generation of gravitational radiation.

\item In the geodesic (non-dissipative) case, the models do not need to be FRW, however  the system relaxes to the FRW spacetime  (if $\Theta > 0$). Such tendency does not appear for dissipative fluids.
\end{itemize}
\subsection{The perfect, geodesic fluid}
In \cite{2} we have considered the case of perfect and gedoesic fluid, without assuming {\it ab initio} the shear--free condition.
As the result of such study we have found that:
\begin{itemize}
\item All possible models compatible with the line element (\ref{1b}) and a perfect fluid, are FRW, and accordingly non--radiating (gravitationally). Both, the geodesic and the non--dissipative, conditions, are quite restrictive, when looking  for a source  of gravitational waves. 

\item  Not only in the case of dust, but also in the absence of dissipation in a perfect fluid, the system is not expected  to radiate (gravitationally) due to the reversibility of the equation of state. Indeed,  radiation is an irreversible process, this fact emerges at once  if  absorption is taken into account and/or Sommerfeld type conditions, which eliminate inward traveling waves, are imposed. Therefore,  the irreversibility of the  process of emission of gravitational waves, must be reflected in the equation of state through an entropy increasing  (dissipative) factor.
\item Geodesic fluids not belonging to the class considered here (Szekeres) have also been shown not to produce gravitational radiation. This strengthens further the case of the non--radiative character of pure dust distributions.
\end{itemize}
\subsection{The dissipative, geodesic fluid}
From the results discussed  above, it becomes clear that  the simplest fluid distribution which we might expect to be  compatible with a gravitational radiation, is a dissipative dust under the geodesic condition. Such a case was analyzed in \cite{4}.

The two  possible subcases were considered separately, namely: the fluid distribution is assumed, from the beginning, to be vorticity--free, or not. 

In the former case, it is shown that the vanishing vorticity implies the vanishing of the heat flux vector, and therefore, as shown in \cite{2}, the resulting spacetime is FRW.

In the latter case, it is shown that the enforcement of the regularity conditions  at the center, implies the vanishing of the dissipative flux, leading also to a FRW spacetime.

Thus  all possible models, sourced by a dissipative  geodesic dust fluid, belonging to the family of the line element considered here, do not radiate gravitational waves during their evolution, unless regularity conditions at the center of the distribution are relaxed.  
Therefore physically acceptable models require the inclusion of, both, dissipative and anisotropic stresses terms, i.e. the geodesic condition must be abandoned. In this case, purely analytical methods are unlikely  to be sufficient to arrive at a full description of the source, and one has to resort to numerical methods.
\section{OPEN ISSUES}
Below we display a partial list of problems which we believe deserve some attention:

\begin{itemize}

\item How could one describe the ``cracking'' (splitting) of the configurations, in the context of this formalism ?
\item We do not have  an exact solution (written down in closed analytical form)  describing gravitational radiation in vacuum, from bounded sources.  Accordingly, any specific modeling of  a source, and its matching to an exterior, should be done numerically. 
\item It should be useful to introduce the concept of the  mass function, similar to the one existing in the spherically symmetric case. This could be relevant, in particular, in the  matching of the source to a specific exterior. 
\item What is the behaviour of the system in the quasi--static approximation? Would be there gravitational radiation in this case?
\end{itemize}

\end{document}